%% file: Kpinunu.tex
\documentclass[10pt,a4paper]{article}
\usepackage[utf8]{inputenc}
\usepackage[english]{babel}
\usepackage{amsmath}
\usepackage{amsfonts}
\usepackage{amssymb}
\usepackage{graphicx}
\usepackage{color}
\definecolor{nicered}{rgb}{0.7,0.1,0.1}
\definecolor{nicegreen}{rgb}{0.1,0.5,.1}

\usepackage[left=2.5cm,right=2.5cm,bottom=5cm]{geometry}

\usepackage{appendix}

\usepackage{geometry}
\usepackage{axodraw}
\usepackage{caption}
\usepackage{subcaption}

\newcommand{\br}[0]{\mathrm{Br}}

\newcommand {\E}[1]{\times 10^{#1}}	

\newcommand{\mrm}[1]{\mathrm{#1}}

\usepackage{mciteplus}

\usepackage{hyperref}
\hypersetup{colorlinks,citecolor=nicegreen,linkcolor=nicered}
\hypersetup{colorlinks=true}

\begin{document}

\begin{center}

{\Large \bf Footprints of leptoquarks: from $ R_{K^{(*)}} $   to $ K \to \pi \nu \bar\nu $ }

\vspace{5mm}

{S. Fajfer$^{a,b}$, N. Ko\v snik$^{a,b}$, L. Vale Silva$^{c}$}

\vspace{3mm}

{\it \small
$^{a}$Jo\v{z}ef Stefan Institute, Jamova 39, P. O. Box 3000, 1001 Ljubljana, Slovenia \\
$^{b}$Faculty of Mathematics and Physics, University of Ljubljana, Jadranska 19, 1000 Ljubljana, Slovenia \\
$^{c}$University of Sussex, Department of Physics and Astronomy, Falmer, Brighton BN1 9QH, UK
}
\vspace{7mm}\\
  \textbf{Abstract}
\end{center}
Rare $K \to \pi \nu \bar \nu$ decays, being
dominated by short distance contributions within the Standard Model (SM), open a window for New Physics  (NP) searches at low energies.   The   $ K \to \pi \nu \bar\nu $ branching ratios are expected to be measured with  $ \sim 10\% $  accuracies by NA62/CERN and KOTO/JPARC. The theoretical uncertainties of branching ratios within the SM are well under control. In the $ B $ sector, it is tentative to explain the $ B $-meson anomalies $ R_{D^{(\ast)}} $ and/or $ R_{K^{(\ast)}} $ by effects of physics beyond the SM.  Although NP seems to be present in the third fermion generation it might also manifest in the flavor changing neutral current transition $ s \to d $. Together with the anticipated good experimental sensitivities and accurate theoretical predictions for $ K\to \pi \nu \bar \nu  $, this motivates studies of correlated effects of NP in rare $ K\to \pi \nu \bar \nu  $ and $ B \to K^{(*)} \mu^+ \mu^- $ decays. Here we consider the loop induced  effects in $ K \to \pi \nu \bar\nu $ in two leptoquark models designed to address lepton-flavor universality violation  in the $  R_{K^{(\ast)}} $ anomalies.

\section{Introduction}
There has been an increasing interest in extensions of the Standard Model (SM) incorporating Lepton Flavor Universality Violation (LFUV), motivated by the measurements of the LFU sensitive observables $ R_{K^{(\ast)}}, R_{D^{(\ast)}} $ in $ B $-meson decays.
Individually, to mention first neutral current processes, the tension in the measurements of $ R_K $ and $ R_{K^\ast} $ (testing lepton universality among muons and electrons), the latter over two distinct di-lepton invariant mass bins, are 2.1--2.6$\sigma$ away from the predictions made in the SM~\cite{Aaij:2014ora,1705.05802}.
The tension is even more significant for charged current processes (testing lepton universality among taus and the light charged leptons), with individual tensions in the range $ \sim (2-3.4) \sigma $, see Refs.~\cite{1205.5442,Lees:2013uzd,Aaij:2015yra,1507.03233,Hirose:2016wfn,Sato:2016svk}.
Very recently, the measurement of the ratio $ R_{J/\psi} $, with a tension with the SM of $ \sim 2\sigma $~\cite{Aaij:2017tyk}, adds a further piece of evidence for LFUV in $b \to c$ transitions, while tensions are also found in branching ratios and angular observables of the $ b \to s \mu \mu $ transition~\cite{Aaij:2014pli,Aaij:2016flj,Aaij:2015esa,Aaij:2015oid,Abdesselam:2016llu,Wehle:2016yoi,ATLAS:2017dlm,CMS:2017ivg}.
Though it is certainly valid to be cautious on interpreting these tensions as true manifestations of New Physics (NP), their good theoretical control (hadronic uncertainties largely cancel out in the ratios) and coherency in terms of a LFUV picture justifies different attempts to extend the SM.\footnote{In the SM, the source of LFUV comes from Yukawa couplings of the SM Higgs to the charged leptons.}

New effects in  flavor changing neutral current  (FCNC) $ b \to s $ and/or  charged current $ b \to c $ semi-leptonic decays are likely to be accompanied by new effects in other quark-flavor transitions. The correlation depends strongly on the specific NP model, and a good candidate for explaining $ R_{K^{(\ast)}} $ and/or $ R_{D^{(\ast)}} $ must satisfy other experimental constraints, such as low-energy flavor data, high-precision electroweak observables, as well as high-energy collider data. More concretely, the authors of \cite{Bordone:2017lsy} have adopted an Effective Field Theory approach assuming that NP couples dominantly to the  third generation. They  focused on the transitions involving $\tau$ leptons and $\tau$ neutrinos and found that branching ratios for $K \to \pi \nu \bar \nu$ could exhibit large deviations from the SM predictions, if the ratios $R_{D^{(*)}}$ get modified by $\sim 20\%$ relative to the SM prediction. Correlations of different flavor sectors, including $K \to \pi \nu \bar\nu$, have also been systematically investigated in Ref.~\cite{Bobeth:2016llm}, for the case where a particular class of dimension six operators is added on top of the SM. Also, the correlation between $\epsilon'/\epsilon$ and rare kaon decays in the context of leptoquark models has been explored recently in Ref.~\cite{Bobeth:2017ecx}.
The authors of  Ref.~\cite{DAmbrosio:2017wis} explored the flavor structure of custodial Randall-Sundrum (RS) models in the light of  observed deviations in the $ B $ decays 
and found that the $K \to \pi \nu \bar \nu$ decays might distinguish among different scenarios  of lepton compositeness.

Motivated by these efforts, we are interested here in FCNC $ s \to d  \nu \bar\nu $ (and $s \to d \ell^+ \ell^-$) transitions.  
In particular, the branching ratio for $ K^\pm \to \pi^\pm \nu \bar\nu $  has been measured with $ \sim 100\% $ uncertainty, while for the $ K_L \to \pi^0 \nu \bar\nu $ decay rate only an upper bound two orders of magnitude above the SM expectation value is known at present.
This situation will improve in the coming years, since the experiments NA62 at CERN and KOTO at JPARC will achieve  experimental accuracies of around ten-percent. On the theoretical front, the uncertainties are well under control due to the absence of long-distance effects, that conversely plague the analogous $ s \to d \ell^+ \ell^- $, or radiative $ s \to d $, transitions (and analogous annihilation topologies).

In this paper, we focus our attention on leptoquark~(LQ) models. More specifically, we consider two distinct LQ attempts to explain LFUV data, whose phenomenological aspects have been discussed recently in the literature: one model where the new contributions to $ b \to s \mu^+ \mu^- $ appear first at the loop-level, and a second LQ model where a contribution to $ b \to s \mu^+ \mu^- $ is present already at tree-level.

The paper is organized as follows. In Section~\ref{sec:differentModels} we introduce the LQ models under consideration and comment on their roles in $B$-meson decays. Then, in Section~\ref{sec:phenoKtopi}, we discuss their contributions to the decays $ K^\pm \to \pi^\pm \nu \bar\nu $ and $ K_L \to \pi^0 \nu \bar\nu $.
In Section~\ref{sec:concl} we conclude with our final comments. In Appendix~\ref{app:numValues} we provide some numerical values and in Appendix~\ref{app:loopfunctions} we discuss technical issues related to the loop-functions, together with a brief discussion of the process $ K^\pm \to \pi^\pm \mu^+ \mu^- $ in the light of the two LQ models.

\section{Framework}\label{sec:differentModels}
The $ B $-meson puzzles have been interpreted by means of an effective Lagrangian approach~\cite{Buttazzo:2017ixm,Alonso:2015sja} and the structure of the four-fermion Lagrangian explaining $ R_{D^{(\ast)}} $ and/or $ R_{K^{(\ast)}} $ anomalies has been well studied. However, it was found in a full-fledged model, matched onto the effective Lagrangian, that it is very difficult to accommodate within $1\sigma$ the large value of $R_{D^{(\ast)}}$~\cite{Dorsner:2017ufx}. On the other hand it seems that $ R_{K^{(\ast)}} $ can be explained by a number of leptoquarks which contribute to $b\to s \mu^+ \mu^-$ transition either at the tree-level~\cite{Buttazzo:2017ixm,Hiller:2014yaa,Becirevic:2015asa,Barbieri:2015yvd,Becirevic:2016oho,Hiller:2016kry,Chen:2017hir,Cheung:2016fjo,Cox:2016epl,Kumar:2016omp,Crivellin:2017zlb,Hiller:2017bzc,Alok:2017sui,Aloni:2017ixa,Calibbi:2017qbu,DiLuzio:2017vat} or at the loop-level~\cite{Bauer:2015knc,Becirevic:2017jtw,Cai:2017wry}.

\subsection{New Physics in $R_{K^{(*)}}$}
Here we briefly present some relevant aspects of rare $ B $-meson decays. Model-independent analyses of the anomalies found in $ B \to K^{(\ast)} \ell^+ \ell^- $, $ \ell = e, \mu $,  and $ B_s \to \phi \mu^+ \mu^- $ point to a possible NP that would result in the left-handed currents operator $ (\bar{s} \gamma_\mu P_L b) \times (\bar{\mu} \gamma^\mu P_L \mu) $. Defining the effective Lagrangian
\begin{equation}
\mathcal{L}^{\rm NP}_{{\rm eff}: b \to s} = \frac{4 G_F}{\sqrt{2}} V_{tb} V^\ast_{ts} \left( \delta C_{9 \mu} Q^{\mu \mu}_9 + \delta C_{10 \mu} Q^{\mu \mu}_{10} \right) + {\rm h.c.}
\end{equation}
with
\begin{equation}
\label{eq:SMopsLchirality}
Q^{\mu \mu}_9 = \frac{\alpha}{4 \pi} (\bar{s} \gamma_\mu P_L b) \times (\bar{\mu} \gamma^\mu \mu) \, , \quad Q^{\mu \mu}_{10} = \frac{\alpha}{4 \pi} (\bar{s} \gamma_\mu P_L b) \times (\bar{\mu} \gamma^\mu \gamma_5 \mu) \, ,
\end{equation}
where $\alpha$ is the fine-structure constant and $V_{ij}$ stands for the CKM matrix element, we have the following favored interval~\cite{Capdevila:2017bsm}\footnote{Note that Ref.~\cite{Becirevic:2017jtw} used slightly different value of $\delta C_{9\mu,10\mu}$.}
\begin{equation}\label{eq:C9NPSeb}
\delta C_{9 \mu} = - \delta C_{10 \mu} = -0.61^{+0.13}_{-0.12} \;\; @ \;\; 1 \sigma, 
\end{equation}
which amounts to a $ \sim 10$--$20\% $ shift of the SM values of the Wilson coefficients $C^{\rm SM}_{9,10}$.\footnote{At the energy scale $ 4.8 $~GeV, $ C^{\rm SM}_{9} \approx 4 $ and $ C^{\rm SM}_{10} \approx -4 $ \cite{DescotesGenon:2012zf}.} To show the correlations of $ K \to \pi \nu \bar\nu $ rates with $ R_K $, we consider the linearized expression~\cite{Hiller:2014ula}
\begin{equation}
R_K = 1 + \frac{2{\rm Re} \{ C^{\rm SM}_{9} (\delta C_{9 \mu})^\ast + C^{\rm SM}_{10} (\delta C_{10 \mu})^\ast \}}{|C^{\rm SM}_{9}|^2 + |C^{\rm SM}_{10}|^2}, 
\end{equation}
valid approximately also for $ R_{K^\ast} $ when operators of flipped quark chirality, compared to Eq.~\eqref{eq:SMopsLchirality}, are not present.

Furthermore, the processes $ B \to K^{(\ast)} \nu \bar\nu $ are expected to give valuable information on the NP interpretation of the aforementioned $ B $-anomalies. In our study the  ratios  $R^{(\ast)}_{\nu \nu}$ are defined as
\begin{equation}
R^{(\ast)}_{\nu \nu} = \frac{\br(B \to K^{(\ast)} \nu \bar\nu)_\mathrm{SM+NP}}{\br(B \to K^{(\ast)} \nu \bar\nu)_\mathrm{SM}}, 
\label{enn}
\end{equation}
subjected to the following experimental bounds for $ K $ and $K^\ast$ channels~\cite{Grygier:2017tzo,Buras:2014fpa} (both at $90\%$~CL)
\begin{equation}
  \label{BKnunu-belle}
\begin{split}
R_{\nu \nu} &< 3.9, \qquad R^{\ast}_{\nu \nu} < 2.7. 
 \end{split}  
\end{equation}


\subsection{Scalar leptoquarks}
Leptoquarks are particularly interesting since they allow interactions between quarks and leptons (see e.g. \cite{Dorsner:2016wpm} for a review). Although we could consider scalar or vector leptoquarks it seems that \textit{light} scalar leptoquarks (with masses of the order of 1~TeV) are simpler to accommodate within  GUT models, while relatively light vector leptoquarks   are difficult to accommodate within any ultra-violet complete theory~\cite{Greljo:2015mma,Fajfer:2015ycq,Buttazzo:2017ixm}. In view of this feature, we prefer not to discuss the phenomenological aspects of vector LQ models regarding $ s \to d \nu \bar\nu $ transitions, but rather note that Refs.~\cite{Barbieri:2015yvd,Barbieri:2016las} discuss some tree-level effects.

According to the classification in~\cite{Dorsner:2016wpm}, scalar leptoquarks which are doublets of $SU(2)_L$ cannot destabilize the proton via the diquark coupling, since $3 B+L =0$, $B$ and $L$ being baryon and lepton numbers\footnote{For $B$ and $L$ violation in the scalar potential, see \cite{Arnold:2012sd,Arnold:2013cva}.}, and therefore can be relatively light. Moreover, it was pointed out in~\cite{Dorsner:2017wwn,Dorsner:2017ufx} that the weak triplet leptoquark, although having $3B +L = 2$, when embedded in a $SU(5)$ GUT model does not destabilize the proton.  The weak doublet leptoquarks may have weak hypercharges $7/6$ or $1/6$. In what follows, different leptoquarks are denoted by their transformation under $(SU(3), SU(2)_L, U(1)_Y)$ and we adopt here the notation introduced in Ref.~\cite{Dorsner:2016wpm}.

\subsubsection{$R_2 (\mathbf{3}, \mathbf{2}, 7/6) $ scalar LQ}
New physics exclusively in semi-leptonic processes naturally evokes LQ frameworks since flavor changing processes involving four-lepton and four-quark transitions are loop suppressed. Despite that, it might seem sensible to invoke a LQ model that contributes to $b \to s \ell \ell $ process at one-loop level (with contributions to $ b \to c \ell \nu $ at tree-level) and thus mimics the SM amplitudes hierarchy, and may therefore be adequate to address $ R_{K^{(\ast)}} $ and $ R_{D^{(\ast)}} $ anomalies simultaneously. An approach along these lines has been followed in~\cite{Bauer:2015knc} but large couplings needed to explain $R_{K^{(*)}}$ at loop-level and $R_{D^{(*)}}$ at tree-level are difficult to reconcile with all available flavor constraints~\cite{1608.07583}. Here we follow the approach of~\cite{Becirevic:2017jtw} where the anomalies in $ b \to s \ell \ell $ decays are accounted for at one-loop level.

The Lagrangian describing the interaction of $R_2(\mathbf{3},\mathbf{2},7/6)$ with quarks and leptons is given by~\cite{Dorsner:2013tla,Dorsner:2016wpm}:
\begin{equation}
  \label{eq:LagDelta}
\begin{split}
\mathcal{L}_{R_2} &= ( V g_R )_{ij} \bar{u}^i P_R e^j R_2^{5/3} + ( g_R )_{ij} \bar{d}^i P_R e^j R_2^{2/3} \\
 &\phantom{=}+ ( g_L )_{ij} \bar{u}^i P_L \nu^j R_2^{2/3} - ( g_L )_{ij} \bar{u}^i P_L e^j R_2^{5/3} + {\rm h.c.} \, ,
\end{split} 
\end{equation}
where $ P_{L (R)} = (1 \mp \gamma_5) / 2 $. The neutrino masses are
negligible in $K$ decays, and then the PMNS matrix reduces to the
identity matrix, $ \mathbf{1}_3 $. In order to explain $ R_{K^{(\ast)}} $ anomalies it was suggested in Ref.~\cite{Becirevic:2017jtw}  that the Yukawa matrices $ g_R $ and $ g_L $ have the following textures
\begin{equation}\label{eq:texturesR2}
g_R = \begin{pmatrix}
0 & 0 & 0 \\
0 & 0 & 0 \\
0 & 0 & ( g_R )_{b \tau} \\
\end{pmatrix} \, , \qquad g_L = \begin{pmatrix}
0 & 0 & 0 \\
0 & ( g_L )_{c \mu} & ( g_L )_{c \tau} \\
0 & ( g_L )_{t \mu} & ( g_L )_{t \tau} \\
\end{pmatrix}
\, .
\end{equation}
Throughout this article, coupling constants are always taken to be real. Regarding their allowed values, the process $ \tau \to \mu \gamma $ receives a LQ contribution proportional to $ ( g_R )_{b \tau} ( g_L )_{t \mu} V_{tb} (m_t / m_\tau) $, which is chiraly enhanced, see \cite{Becirevic:2017jtw}. In order to allow for a large $ ( g_L )_{t \mu} $, $ ( g_R )_{b \tau} $ must be suppressed and here we set it to zero. Therefore, the only LQ couplings to fermions are given by $ g_L $. Moreover, the couplings $ g^{c \tau}_L $ and $ g^{t \tau}_L  $ are strongly constrained by the same process $ \tau \to \mu \gamma $ when $ g^{c \mu}_L, g^{t \mu}_L $ are both large, of order $ \mathcal{O} (1) $.

With the ansatz for $g_L$ in Eq.~\eqref{eq:texturesR2} and the aforementioned constraints, tree-level contributions to $ b \to c \ell \nu $ are absent. Note, however, that a different coupling texture was used in Ref.~\cite{Dorsner:2013tla,Freytsis:2015qca} to explain the $ R_{D^{(\ast)}} $ anomalies. See also~\cite{Chauhan:2017ndd} for a different texture of LQ Yukawa couplings.


\begin{figure}
\begin{center}
\includegraphics[scale=0.55]{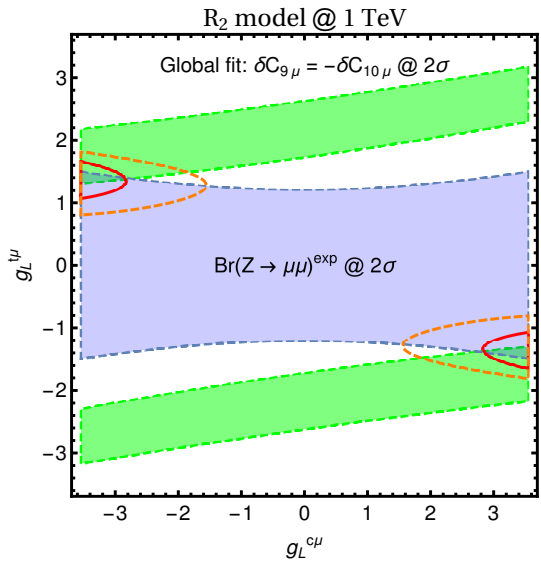}
\end{center}
\caption{\it Contraints in the plane $ \{ g^{c \mu}_L , \, g^{t \mu}_L \} $: in purple we show the region allowed by $ \br (Z \to \mu \mu)^{\rm exp} $; in green, the bound from Eq.~\eqref{eq:C9NPSeb} below. The solid red line (dahed orange line) delimits the $ 1 \sigma $ (respectively, $ 2 \sigma $) combined region. Here, $ m_{R_2} = 1 $~TeV. (Colors online.)}\label{fig:figZmumu}
\end{figure}

In order to further constrain this model, the following bounds are also important:
\begin{itemize}
\item[(a)] in order to explain the $ R_{K^{(\ast)}}$ anomalies  we are compelled to adjust the NP Wilson coefficient $ \delta C_{9 \mu} = - \delta C_{10 \mu} $,
\item[(b)] the constraint coming from the difference between the experimental and theoretical  esults for the muon anomalous magnetic moment  $a_\mu^\mathrm{exp}-a_\mu^\mathrm{SM}  =\Delta a_\mu = (2.88 \pm 0.63 \pm 0.49)\times 10^{-9}  $~\cite{Jegerlehner:2009ry,NYFFELER:2014pta} (see, e.g., \cite{Patrignani:2016xqp} for the breakdown of uncertainties),
\item[(c)] the precise experimental value of $ \br (Z \to \mu \mu) $, measured at LEP.
\end{itemize}
These bounds and requirements are summarized in Figure~\ref{fig:figZmumu}. Note that the $\br(Z \to \mu\mu)$ is sensitive to large values of $ g_L^{t \mu} $. In this figure, we limit the couplings to $ |g_L^{c \mu}|^2, |g_L^{t \mu}|^2 \leq 4\pi $ in order to stay within the perturbative regime. For the combined $ 1\sigma $ regions indicated, the predicted anomalous magnetic moment of the muon gets further worsened by $ \gtrsim 1\sigma $. We have also checked that charmonia decays~\cite{Hazard:2016fnc,Aloni:2017eny} do not impose important bounds. The expression for $R_{\nu\nu}^{(*)}$ can be related to loop-induced amplitudes of $ b \to s \mu \mu $ transitions, however the resulting constraints are weak.

Note that LHC constraints of flavored processes at high energies are becoming sensitive to flavor couplings employed in low-energy flavor phenomenology, see, e.g.,~\cite{Greljo:2017vvb}. Their analysis sets constraints on effective four-fermion operators contributing to $ p p \to \mu^+ \mu^- $ at the tail of the di-lepton invariant mass spectrum. Therefore, the results apply directly for a NP spectrum beyond $ \rm{few} $~TeV, but they may still remain indicative at the region $ 1 $~TeV, in which case a rough estimate of $g_L^{c \mu}$ from their study gives $ | g_L^{c \mu} | \lesssim 0.8 $. This is not very different from the values used in  our analysis, but for a more precise knowledge of this coupling, dedicated analyses of LHC data would clarify the situation.

\subsubsection{$S_3 (\mathbf{\bar{3}}, \mathbf{3}, 1/3)$ scalar LQ}\label{sec:constraintsTripletLQ}
We now discuss a different mechanism for explaining the anomalies in $ b \to s \ell \ell $ data. In Ref.~\cite{Dorsner:2017ufx,Dorsner:2017wwn}, a model with two LQs has been considered, namely $ S_3 = (\mathbf{\bar{3}}, \mathbf{3}, 1/3) $ and $ \tilde{R}_2 = (\mathbf{3}, \mathbf{2}, 1/6) $. The authors of that model noticed that these two leptoquark states are important for the one-loop neutrino mass mechanism within the framework of grand unification theory~(GUT)~\cite{Dorsner:2017ufx,Dorsner:2017wwn}.  
In Ref.~\cite{Dorsner:2017ufx} an attempt was done to explain all $ B $-meson anomalies in this GUT setup, accounting for all the existing low-energy constraints. As already noticed by the authors of~\cite{Buras:2014fpa,Crivellin:2017zlb} the current bounds on $R_{\nu\bar\nu}^{(*)}$ are rather restrictive for the leptoquark models and since $S_3$ and $\tilde R_2$ contribute to $B \to K^{(*)} \nu \bar \nu$ at tree-level they cannot fully accommodate the $R_{D^{(*)}}$ anomaly. The role of $\tilde R_2$ in LFU anomalies in this setting has been shown to be minor and thus we omit the $\tilde R_2$ state from further discussion and focus on scenarios with light $S_3$ as studied in~\cite{Dorsner:2017ufx}.

Following the notation of~\cite{Dorsner:2017ufx},  the  Lagrangian describing the interactions of  the weak-isospin triplet $ S_3 $ and  the SM fermions is \cite{Dorsner:2016wpm}
\begin{equation}\label{eq:LagS3}
  \begin{split}
     \mathcal{L}_{S_3} &= - y_{ij} \bar{d}^C_L {}^i \nu^j_L S^{1/3}_3 - ( V^\ast y )_{ij} \bar{u}^C_L {}^i e^j_L S^{1/3}_3 \\ 
&\phantom{=}- \sqrt{2} y_{ij} \bar{d}^C_L {}^i e^j_L S^{4/3}_3 + \sqrt{2} ( V^\ast y )_{ij} \bar{u}^C_L {}^i \nu^j_L S^{-2/3}_3 + {\rm h.c.}
  \end{split}
\end{equation}
In order to avoid all existing experimental constraints on the first down-type quark and charged lepton generation at tree-level, the Yukawa matrix $ y $ assumes the following texture~\cite{Dorsner:2017ufx}, 
\begin{equation}
y = \begin{pmatrix}
0 & 0 & 0 \\
0 & y_{s \mu} & y_{s \tau} \\
0 & y_{b \mu} & y_{b \tau} \\
\end{pmatrix}.
\end{equation}
We will study the following two scenarios for the LQ couplings:
\begin{eqnarray}
\textrm{Scenario I}:&&  \{ y_{s\mu}, y_{b\mu} \} \; \textrm{free parameters} \, , \; y_{s\tau}=y_{b\tau} = 0, \; \nonumber\\
\textrm{Scenario II}:&&  \{ y_{s\mu}, y_{b\mu}, y_{s\tau}, y_{b\tau} \} \; \textrm{free parameters} \, . \nonumber 
\end{eqnarray}
These correspond to the scenarios studied in Sections 5.1 and 5.2 of~\cite{Dorsner:2017ufx}. Under the constraints discussed in \cite{Dorsner:2017ufx}, the best fit points, for the leptoquark mass set to $m_{S_3} = 1$~TeV, are
\begin{eqnarray}
\textrm{Scenario I}: && y_{s\mu} = -0.002 \, , \; y_{b\mu} = -0.46 , \nonumber\\
\textrm{Scenario II}: &&  y_{s\mu} = -0.047 \, , \; y_{b\mu} = -0.020 \, , \; y_{s\tau} = 0.87 \, , \; y_{b\tau} = -0.048. \nonumber
\end{eqnarray}
Solutions with overall sign flips lead to the same results. In the two scenarios presented above, the SM hypothesis, namely, $ \{ y_{s\mu}, y_{b\mu}, y_{s\tau}, y_{b\tau} \} \, = 0 $, has a similar pull with respect to the hypotheses described above. In our predictions we will use the $1\sigma$ regions of Yukawa couplings presented in~\cite{Dorsner:2017ufx}.

\section{Leptoquarks in  rare $ K $ decays}
\label{sec:phenoKtopi}
The SM predictions of rare $ K \to \pi \nu \bar\nu $ decays have achieved a great level of precision and robustness, see \cite{Buchalla:1993wq,Misiak:1999yg,Gorbahn:2004my,Isidori:2005xm,Buras:2005gr,Buras:2006gb,Mescia:2007kn,Brod:2008ss,Brod:2010hi}. 
The effective Lagrangian describing the SM short-distance transition $s \to d\nu\bar\nu$ is given by
\begin{equation}
{\cal L}^{\rm SM}_{{\rm eff}: s \to d} = \frac{4 G_F}{\sqrt{2}}\frac{\alpha}{2\pi} V_{ts}^*V_{td} C^{{\rm SM}}_{sd,\ell} \left(\bar s \gamma_\mu P_L d \right) \times \left(\bar\nu_{\ell}\gamma^\mu P_L \nu_{\ell}\right) + {\rm h.c.}. \label{Leff}
\end{equation}
The Wilson coefficient in the SM reads
\begin{equation}
C^{{\rm SM}}_{sd,\ell} = -\frac{1}{s_w^2}\left( X_t + \frac{V_{cs}^*V_{cd}}{V_{ts}^*V_{td}} X_c^\ell\right),\label{CSM}
\end{equation}
where $X_t$ and  $X_c^\ell$ denote the loop-functions for the top and charm contributions, respectively, and $s_w$ is the sine of the weak mixing angle. In agreement with the inputs in Appendix~\ref{app:numValues}, the value of the loop-function $ X_t $ is $ X_t \simeq 1.506 $, and includes NLO QCD corrections~\cite{Buchalla:1993wq,Misiak:1999yg} as well as NLO EW corrections~\cite{Brod:2010hi}. The loop-functions $X_c^\ell$ are known up to NNLO QCD corrections~\cite{Gorbahn:2004my,Buras:2005gr,Buras:2006gb} and NLO EW corrections~\cite{Brod:2008ss}.

Following \cite{Buchalla:1998ba} the branching ratio for $K^+\to \pi^+\nu\bar\nu$ in the SM can be obtained after summation over the three neutrino flavor contributions 
\begin{equation}
\br(K^+\to\pi^+\nu\bar\nu)_{\rm SM} = C_K \sum_{\ell = e,\mu,\tau} \left| \frac{V_{ts}^*V_{td}}{\lambda^5}X_t + \frac{V_{cs}^*V_{cd}}{\lambda}\left(\frac{X_c^\ell}{\lambda^4} + \delta P_{c,u} \right)\right|^2,\label{Kpvv}
\end{equation}
where $C_K= {\kappa_+ (1+\delta_{\rm em})}/{3}$, with $\kappa_+ = (5.173\pm 0.025)\times 10^{-11}(\lambda/0.225)^8$, see Ref.~\cite{Mescia:2007kn}, $\delta_{\rm em} = -0.003$ is a QED correction,
and $\delta P_{c,u} \approx 0.04\pm 0.02$ is the long-distance contribution from light-quark loops~\cite{Isidori:2005xm} (see also~\cite{Bai:2017fkh} for an exploratory lattice calculation).
The decay $K_L\to \pi^0\nu\bar\nu$ in the SM is CP violating and lepton-flavor universal:
\begin{equation}
\br(K_L\to\pi^0\nu\bar\nu)_{\rm SM} = \kappa_L \left| \frac{{\rm Im}[ V_{ts}^*V_{td} ]}{\lambda^5} X_t \right|^2,
\end{equation}
\noindent
where $ \kappa_L = (2.231 \pm 0.0013) \times 10^{-10} (\lambda/0.225)^8$, see \cite{Mescia:2007kn}. The SM predictions for these two branching ratios are
\begin{equation}
  \label{eq:predSM}
  \begin{split}
    &\br (K^+ \to \pi^+ \nu \bar\nu)_\mathrm{SM} = 0.882^{+0.092}_{-0.098} \times 10^{-10} ,\\
&\br (K_L \to \pi^0 \nu \bar\nu)_\mathrm{SM} = 0.314^{+0.017}_{-0.018} \times 10^{-10},
  \end{split}
\end{equation}
which are the results from~\cite{JOcariz}, including tree- and loop-level dominated observables in the extraction of the CKM matrix elements in the SM (as opposed to Ref.~\cite{Buras:2015qea}), with a larger uncertainty in the charged mode due to the light quark contributions. On the other hand, the experimental values are~\cite{Adler:2001xv,Anisimovsky:2004hr,Artamonov:2008qb,Artamonov:2009sz,Ahn:2009gb}
\begin{equation}
  \begin{split}
&\br (K^+ \to \pi^+ \nu \bar\nu)^\mathrm{exp} < 3.35 \times 10^{-10} \;\; @ \;\; 90\% \; {\rm CL},\\ 
&\br (K_L \to \pi^0 \nu \bar\nu)^\mathrm{exp} < 2.6 \times 10^{-8} \;\; @ \;\; 90\% \; {\rm CL} .
  \end{split}
\end{equation}
As mentioned above, an anticipated precision of $ 10\% $, or even better in the long term, for both channels is expected for NA62 and KOTO~\cite{Pepe:2015kaa,Shiomi:2014sfa}.

\subsection{$R_2 (\mathbf{3}, \mathbf{2}, 7/6) $ scalar LQ in {$K \to \pi \nu \bar \nu$}} \label{sec:R2rarekaonpheno}

\begin{figure}
\begin{center}
\includegraphics[scale=0.5]{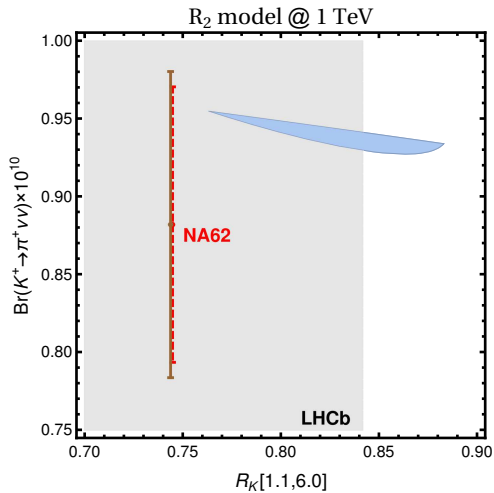}
\end{center}
\caption{\it Correlation between $ R_K [1.1, 6.0] $ (where the values in brackets give the interval of the di-lepton invariant mass bin) and $ \br(K^+ \to \pi^+ \nu \bar\nu) $. The LHCb measurement of $ R_K $ is shown in grey. The expected future experimental accuracies for the rate of $ K^+ \to \pi^+ \nu \bar\nu $ is shown in red, dashed line, with central value given by the theoretical prediction under the SM. In blue the $ 1 \sigma $ region satisfying the contraints (a)-(c) described in the main text when varying $ \{ g^{c \mu}_L , \, g^{t \mu}_L \} $. We stress that the blue region does not include the uncertainties of the SM theoretical predictions, indicated by the solid brown line. $ m_{R_2} = 1 $~TeV. (Colors online.)}\label{fig:fig1}
\end{figure}

The $s \to d \nu \bar \nu$ process is mediated in this scenario via a box diagram similar to the one shown in Fig.~\ref{fig:sub4}. 
The corresponding loop-function can be found in~\cite{Becirevic:2017jtw}. In our analysis we will keep only the interference term of LQ with the SM amplitude.

We now discuss the resummation of large factors $ \alpha_s  \log x_c $ by the use of renormalization group equations in the leading log approximation. We first consider the contribution proportional to $ \lambda_c = V_{cd} V^\ast_{cs} $, resulting from the box with two charm quarks (and also diagrams with the up quark). After the top, the $ W $ boson and $ R_2 $ have been integrated out at the energy scale $ \mu_{tW} = \mathcal{O} (m_t, M_W, m_{R_2}) $, one is left with the operator structures $ (\bar{s} \gamma^\mu P_L q') \times (\bar{q} \gamma_\mu P_L d) $, $ (\bar{q}' \gamma^\mu P_L q) \times (\bar{\nu}_\ell \gamma_\mu P_L \nu_\ell) $, $ q, q' = u, c $, and $ (\bar{s} \gamma^\mu P_L d) \times (\bar{\nu}_\ell \gamma_\mu P_L \nu_\ell) $, similarly to the SM case after the top, the $ W $ and the $ Z $ bosons have been integrated out from the full theory. The following factor summarizes short-distance QCD corrections to the charm box contribution (after neglecting the bottom quark threshold) taking leading order~(LO) anomalous dimensions from  \cite{Buchalla:1993wq}
\begin{equation}\label{eq:etafactor}
\bar{\eta}^{(R_2)}_{cc} (\mu_{tW}) \equiv \frac{1}{\log (x_c (\mu_c))} \frac{2 \pi}{\alpha_s (\mu_c)} \sum_{i = \pm} \frac{x_i}{x - \gamma_i} \left[ \left( \frac{\alpha_s (\mu_{tW})}{\alpha_s (\mu_c)} \right)^{d} - \left( \frac{\alpha_s (\mu_{tW})}{\alpha_s (\mu_c)} \right)^{d_i} \right] \, ,
\end{equation}
with
\begin{equation}
\begin{split}
 \gamma_\pm &= \pm 6 \, (N_c \mp 1) / N_c \, , \quad x_\pm = 8 (1 \pm N_c) \, , \quad \gamma_m = 3 (N_c^2 - 1) / N_c \, , \\
 x &= 2 \, \left( \gamma_m - \beta^{(N_f)}_0 \right) \, , \quad d = x / (2 \beta^{(5)}_0) \, , \quad d_i = \gamma_i / (2 \beta^{(5)}_0) \, , 
\end{split}  
\end{equation}
where $ N_c $ is the number of colors, and $ \beta^{(N_f)}_0 = (11 N_c - 2 N_f) / 3 $, with $ N_f $ the number of dynamical flavors.
For $ \mu_{tW} = M_W $, we then have $ \bar{\eta}^{(R_2)}_{cc} (M_W) = 0.8 $, therefore only slightly damping the charm box contribution, where the following numerical values have been employed: $ \alpha_s (M_Z) = 0.1185 $, $ \Lambda^{(4)}_{\overline{\rm MS}} = 0.327 $ and $ \mu_c = 1.3~{\rm GeV} $.\footnote{Although the LO anomalous dimension matrix has been considered, we use the expression of the strong coupling constant up to the Next-to-Leading Order (NLO).} As usual, there is a large dependence of the LO value of $ \bar{\eta}^{(R_2)}_{cc} (\mu_{tW}) $ on the value of $ \mu_{tW} $: varying $ \mu_{tW} $ over the interval $ [M_W / 2 , 1~{\rm TeV}] $, results in $ \bar{\eta}^{(R_2)}_{cc} (\mu_{tW}) $ in the range $ [0.4, 0.9] $, where the smaller value corresponds to $ \bar{\eta}^{(R_2)}_{cc} (1~{\rm TeV}) $.
The situation is simpler in the cases of the contributions proportional to $ \lambda_t = V_{td} V^\ast_{ts} $, resulting from the box with two top quarks, and $ V^\ast_{ts} V_{cd} $ or $ V^\ast_{cs} V_{td} $, resulting from the boxes with top and charm quarks, since there only the operator $ (\bar{s} \gamma^\mu P_L d) \times (\bar{\nu}_\ell \gamma_\mu P_L \nu_\ell) $ is required. Note that for $ K_L \to \pi^0 \nu \bar{\nu} $, the top contribution is the only one relevant; for different reasons, the top contribution is largely dominant for the transition $ b \to s \ell^+ \ell^- $.

A few further precisions are required. To calculate the interference of the LQ contributions with the SM, we employ the values $ X^e_c = (11.31 \pm 0.74) \times 10^{-4} $ and $ X^\tau_c = (7.77 \pm 0.62) \times 10^{-4} $~\cite{Buchalla:1998ba} (with uncertainties estimated from a very conservative uncertainty for the mass of the charm-quark), and $ \delta P_{c, u} \approx 0.04 \pm 0.02 $. Varying these values over their quoted uncertainty ranges, and the ones provided in Appendix~\ref{app:numValues}, we obtain modulations of the relative LQ contributions below $ 1\% $, that we neglect in our calculations. Furthermore, to make the LQ effects transparent in the plots that will be shown, we do not include the SM uncertainties in Eq.~\eqref{eq:predSM}, which are $ \sim 10\% $ for the charged channel and $ \sim 5\% $ for the neutral mode.

In Fig.~\ref{fig:fig1} we show the correlation between $ R_K [1.1, 6.0] $ and $ \br ( K^+ \to \pi^+ \nu \bar{\nu} ) $, which is due to the common couplings $ g_L^{c \mu} $ and $ g_L^{t \mu} $. With respect to the SM values, the highest variations are $ 8\% $ and $ 5\% $ for $ K^+ \to \pi^+ \nu \bar{\nu} $ and $ K_L \to \pi^0 \nu \bar{\nu} $, respectively. Compared to the analogous contribution accommodating the $b \to s \ell^+ \ell^- $ anomalies, the effects in $K \to \pi \nu\bar\nu$ are smaller due to different CKM factors. In particular, the contributions to the charged mode proportional to the large LQ coupling to the charm do not overcome the overall suppression of the LQ mass. Compared to the analogous contribution accommodating the $b \to s \ell^+ \ell^- $ anomalies, the effects in $K \to \pi \nu\bar\nu$ are slightly smaller for the charged channel, but of similar relative size.

\subsection{$S_3(\mathbf{\bar{3}}, \mathbf{3}, 1/3) $ scalar LQ in $K \to \pi \nu \bar \nu$}
\subsubsection{Tree-level effects}\label{sec:treeeffects}
In this Section we make an estimate of the tree-level effects of
$S_3$. In the model considered in~\cite{Dorsner:2017ufx} the $S_3$
Yukawa couplings to the down-quark are set to zero at tree-level,
precisely in order to avoid constraints from kaon decays. Allowing
$ y_{d \mu} $ or $y_{d\tau}$ to be finite, tree-level contributions of
$ S_3 $ to LFV $s\to d \tau \mu$ currents are possible, since the $B$
anomalies explanation require finite $ y_{s \mu}$ and
$y_{s\tau}$. This would then make the lepton flavor violating decay
$ \tau \to \mu K$ an important constraint. The experimental bound for
$ \tau \to \mu K_S^0$~\cite{Carpentier:2010ue,Patrignani:2016xqp}
implies
\begin{equation}
  \label{eq:tauLFV}
|y_{d\tau} y_{s\mu}|,  |y_{s\tau} y_{d\mu}| \lesssim 0.014\,(m_{S_3}/\mrm{TeV})^2  .
\end{equation}
Similarly, the charged current process $\pi^- \to \mu \bar\nu$ depends on possible non-zero Yukawa couplings to the down quark, $y_{d\mu}$, $y_{d\tau}$. The effect on the decay width, normalized to the SM value, reads
\begin{equation}
  \begin{split}   
  \frac{\Delta \Gamma(\pi \to \mu \bar \nu)}{\Gamma(\pi \to \mu \bar \nu)_\mrm{SM}}
  = \frac{-2v^2 y_{d\mu}}{4m_S^2} \,\mrm{Re} &\left[y_{d\mu}+(V_{us}^*/V_{ud}) y_{s\mu}
    + (V_{ub}^*/V_{ud}) y_{b\mu}\right] \\
  &+ \left(\frac{v^2}{4m_S^2}\right)^2 \left|y_{d\mu} + (V_{us}^*/V_{ud}) y_{s\mu} + (V_{ub}^*/V_{ud}) y_{b\mu} \right|^2 \left(|y_{d\mu}|^2 + |y_{d\tau}|^2\right).
  \end{split}
\end{equation}
Experimental relative precision reaches $4\E{-7}$ and allows to put an asymmetric constraint $-18\E{-3} < y_{d\mu} < 3\E{-3}$, if we assume that Yukawas $y_{s\mu}$ and $y_{b\mu}$ are within the $1\sigma$ region of Scenario I~\cite{Dorsner:2017ufx}. More importantly for the $s \to d \nu \bar \nu$ processes one can also extract, in Scenario I, the allowed range  $|y_{d\mu} y_{s\mu}|  \lesssim 10^{-2}$, which directly enters $s \to d \nu \bar\nu$.

In Scenario II, non-zero $y_{d \tau}$ could potentially induce $s \to d \nu \bar \nu$ via $y_{s\tau} y_{d\tau}$. In this case, $y_{d\tau}$ does not interfere with the SM contribution to $\pi \to \mu \bar\nu$ and therefore the bound on $y_{d\tau} y_{s\tau}$ is of order $1$ in the $1\sigma$ parameter space of Scenario II. Much better sensitivity is expected in $s \to d \nu\bar\nu$ processes where this term does interfere with the SM amplitude.


For such a small coupling, effects induced in the up-type sector by $ y_{d \mu} \neq 0 $ (cf. Eq.~\eqref{eq:LagS3}) are negligible and the analysis in~\cite{Dorsner:2017ufx} is not modified. Moreover, new contributions to atomic parity violation, and corrections to meson-mixing in the system of kaons~\cite{Isidori:2010kg} are also negligible.
Further note that since we do not allow for couplings to electrons, $ e - \mu $ conversion receives no contribution from $ S_3 $ exchanges up to one-loop corrections.

For very small values of $ y_{d \mu} $, as given by the above discussion, radiative effects due to weak interaction and proportional to other LQ couplings may become competitive. We now discuss the size of these weak interaction radiative corrections, and after that we discuss their phenomenological impact on $ s \to d \nu \bar\nu $ transitions.

\subsubsection{Radiative Yukawas and boxes}

\input{Diagrams.tex}

The relevant diagrams for the one-loop radiative corrections to $ s \to d \nu \bar\nu $ transitions are shown in Fig.~\ref{fig:diagrams} and include: (SE) a reducible diagram where a $ W $ boson changes flavor of the down-type quark external leg, (V) a vertex correction to the coupling of $ S^{1/3}_3 $ where a $ W $ is exchanged, and (VT) a second vertex correction to $ S^{1/3}_3 $ where a coupling to the $ W $ gauge boson,
\begin{equation}
  \begin{split}
& \mathcal{L}^\mathrm{gauge}_{S_3} \supset + i g \Big( - \partial^\mu S^{2/3}_3 W^-_\mu S^{1/3}_3 + \partial^\mu S^{1/3}_3 W^-_\mu S^{2/3}_3 \\
& \qquad\qquad\qquad\qquad - \partial^\mu S^{4/3}_3 W^-_\mu S^{-1/3}_3 + \partial^\mu S^{-1/3}_3 W^-_\mu S^{4/3}_3 \Big) + {\rm h.c.} \, ,    
  \end{split}
\end{equation}
is present. Apart from this set of topologies, there is a box diagram ``Box" where a $ W $ and a $ S^{2/3}_3 $ LQ are exchanged. Note that the crossed box, (Box'), is only possible for charged leptons in the final state, i.e., in the process $s\to d \ell \ell$.

It is instructive at this point to have a look at the flavor structure of the different contributions. While the standard model (SM), (SE), (V) and (VT) contributions are proportional to the product of two CKM matrix elements, the NP box diagram is proportional to the product of  four CKM matrix elements. We indicate in Table~\ref{tab:flavornunu} the pattern of $ \lambda $ suppressions, together with $ g $ and powers of $ y_{ij} $ ($ i=s,b $ and $ j=\mu,\tau $). It is interesting to note that (SE), (V) and (VT) with an internal top have a relative factor compared to the SM top contribution of $ y_{s \tau}  y_{b \tau}/ (g^2 \lambda^2)$. Together with the loop-functions, possibly resulting in logarithmic enhancements, this sets up the hierarchy of NP contributions.\footnote{We note that the same $ 1/\lambda^2$ enhancements with respect to the SM are not possible for the processes $ B \to K^{(\ast)} \ell^+ \ell^- $ for $ S_3 $, nor for $ {R_2} $: in both LQ scenarios, the NP contributions to $b\to s \nu\bar \nu$ follow the $ V_{tb} V^\ast_{ts} $ structure of the SM.}

\begin{table}[t]
\begin{center}
\begin{tabular}{|c|c|c|c|c|c|}
\cline{2-6}
\multicolumn{1}{c|}{} & \textit{c} & \textit{t} & \textit{cc} & \textit{ct}, \textit{tc} & \textit{tt} \\
\cline{2-6}
\hline
(SM) & $ g^4  \lambda $ & $ g^4  \lambda^5 $ & & & \\
\hline
(SE), (V), (VT) & $ g^2  \lambda  y_{s \tau}^2 $ & $ g^2  \lambda^3  y_{s \tau}  y_{b \tau} $ &&& \\
\hline
(Box) &&& $ g^2  \lambda  y_{s \tau}^2 $ & $ g^2  \lambda^3  y_{s \tau}  y_{b \tau} $ & $ g^2  \lambda^5  y_{b \tau}^2 $ \\
\hline
\end{tabular}
\end{center}
\caption{\it Flavor and gauge coupling factors present in the calculation of the LQ contributions to $ K \to \pi \nu \bar\nu $ compared to the SM, with internal quark-flavors given by the columns and loop-topologies by the lines. The external flavors of the neutrinos are dominantly $ \nu_\tau $ (in Scenario II discussed in the text). The unitarity of the CKM matrix has been employed to include the contributions of the up-quark into those of the charm- and top-quarks. We assume $ y_{s \tau} \gg \lambda^2  y_{b \tau} $ and $ y_{b \tau} \gg \lambda^2  y_{s \tau} $. The gauge coupling constant of the gauge group $ SU(2)_L $ is denoted by $ g $. (For the flavor factors appearing in the calculation of the LQ contributions to $ K \to \pi \mu^+ \mu^- $, one replaces the tauonic flavor for the muonic one.)}\label{tab:flavornunu}
\end{table}

\begin{figure}
\begin{center}
\includegraphics[scale=0.65]{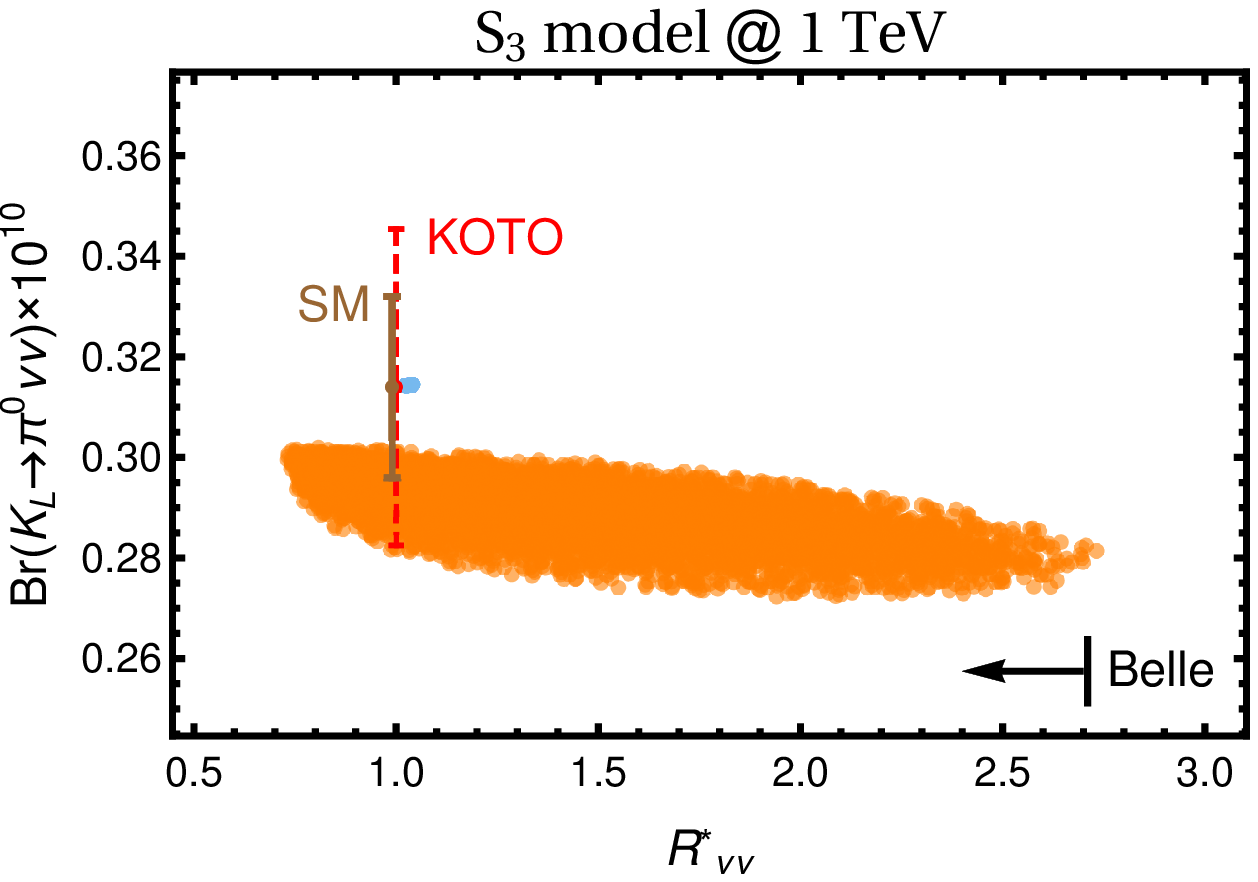}
\end{center}
\caption{\it Correlation between $ R^\ast_{\nu \nu} $, defined in Eq.~\eqref{enn}, and $ \br(K_L \to \pi^0 \nu \bar\nu) $ when there is no tree-level contribution to rare kaon decays in the $ S_3 $ model. The present experimental upper limit by Belle is indicated by the arrow. The expected future experimental sensitivity at KOTO for the rate of $ K_L \to \pi^0 \nu \bar\nu $ is shown in red, dashed line, with central value given by the theoretical prediction within the SM. Blue dot close to the SM central value corresponds to Scenario I. In orange we have the $ 1 \sigma $ region around Scenario II (mirror solutions with an overall sign flip do not change the results). We stress that the presented $1\sigma$ region does not include the uncertainties of the theoretical predictions. Error of the SM theoretical prediction in $K_L \to \pi^0 \nu\bar\nu$ is shown by the brown bar. We consider here $ m_{S_3} = 1 $~TeV. (Colors online.)}\label{fig:fig4_DFFK}
\end{figure}

We relegate the discussion of the calculation of different
diagrams to Appendix~\ref{app:loopfunctions}, and now start with the
discussion of their results.  At the current status of theoretical and
experimental precision, it is sufficient to keep only the top
contributions in the vertex and self-energy topologies, and neglect
the charm contributions and the box topology. In this case,
$ (\bar{s} \gamma^\mu P_L d) \times (\bar{\nu}_\ell \gamma_\mu P_L
\nu_\ell) $ is the only relevant operator structure at the LO, and
thus there is no further corrective factor as in
Section~\ref{sec:R2rarekaonpheno} (cf. Eq.~\eqref{eq:etafactor}). We
then get
\begin{eqnarray}
{\cal L}^{\rm SM + NP}_{\rm eff} (s \to d \nu \bar \nu) &=& \frac{4 G_F}{\sqrt{2}}\frac{\alpha}{2\pi} \left[ V_{ts}^*V_{td} C^{{\rm SM}}_{sd,\ell} + \frac{M_W^2}{M^2_{S_3}} \frac{2\pi}{g^2 \alpha} y^{(0)}_{s \ell} \left( y^{(0)}_{d \ell} - \Gamma^{\rm ren}_\ell (0; M^2_{S_3}) \right) \right] \left(\bar s \gamma_\mu P_L d \right) \left(\bar\nu_{\ell}\gamma^\mu P_L \nu_{\ell}\right) \nonumber\\
&&+   \frac{y^{(0)}_{s i}}{2 M^2_{S_3}} \left( y^{(0)}_{d j} - \Gamma^{\rm ren}_j (0; M^2_{S_3}) \right) \left(\bar s \gamma_\mu P_L d \right) \left(\bar\nu_{i}\gamma^\mu P_L \nu_{j}\right) + {\rm h.c.}
\end{eqnarray}\label{LagrangianSMLQM}
where the superscript ``$ (0) $'' indicates bare couplings, and the calculation of the function $ \Gamma^{\rm ren} $ is discussed in the Appendix~\ref{app:loopfunctions}, with its first argument vanishing when the external momenta are neglected and its second argument indicating the subtraction point in the renormalization procedure. Numerically
\begin{equation}
\Gamma^{\rm ren}_j (0; 1 \; {\rm TeV}^2) = - [ (3.8 + 1.5 i) y_{b j} - (0.16 + 0.06 i) y_{s j} ] \times 10^{-4} \, .
\end{equation}
To indicate more clearly the effects of the radiative corrections, we set $ y^{(0)}_{d \ell} $ to zero. For the process $ K^- \to \pi^- \nu \bar{\nu} $, it then follows
\begin{equation}
  \label{eq:resultKpm}
\begin{split}
 \frac{\br (K^- \to \pi^- \nu \bar\nu)}{\br (K^- \to \pi^- \nu \bar\nu)_\mrm{SM}} \simeq 1& + 1.03 \,y_{b \mu} y_{s \mu} + 0.94 \,y_{b \tau} y_{s \tau} \\
&+ 0.75 \,(y_{b \mu}^2 y_{s \mu}^2 + y_{b \tau}^2 y_{s \tau}^2 + y_{b \mu}^2 y_{s \tau}^2 + y_{b \tau}^2 y_{s \mu}^2) \, , \qquad @ \; M_{S_3} = 1 \; {\rm TeV} \, ,
\end{split}
\end{equation}

\noindent
where the difference between the tau and the muon contributions in the first line results from $ X_c^\tau \neq X_c^\mu $ (see comments in Section~\ref{sec:R2rarekaonpheno}), and for the process $ K_L \to \pi^0 \nu \bar{\nu} $ we have\footnote{Note that complex phases in the CKM matrix are \textbf{CP}-odd while those from the loop-functions are \textbf{CP}-even.}
\begin{equation}
  \label{eq:resultKL}
  \begin{split}
    \frac{\br (K_L \to \pi^0 \nu \bar\nu)}{\br (K_L \to \pi^0 \nu \bar\nu)_\mrm{SM}} \simeq 1& + 1.37 \, (y_{b \mu} y_{s \mu} + y_{b \tau} y_{s \tau}) \\
& + 1.4 \,(y_{b \mu}^2 y_{s \mu}^2 + y_{b \tau}^2 y_{s \tau}^2 + y_{b \mu}^2 y_{s \tau}^2 + y_{b \tau}^2 y_{s \mu}^2) \, , \qquad @ \;M_{S_3} = 1 \; {\rm TeV} \, ,
\end{split}
\end{equation}
where we omit sub-leading contributions in Eqs.~\eqref{eq:resultKpm} and \eqref{eq:resultKL}, and superscripts ``$ (0) $'' have been dropped in these two equations for readability. The above numerical values are also enhanced by a constructive interference among the two vertex topologies (in the unitary gauge). Despite many enhancements, resulting in the large numerical pre-factors seen above, the net effect is of order $10\%$ for both modes, due to the strong constraints the different LQ couplings are subjected to, cf. Section~\ref{sec:constraintsTripletLQ}. This shows the importance of the phenomenological analysis of~\cite{Dorsner:2017ufx} on studying the effects of $ S_3 $ in rare kaon decays. This is illustrated in Figure~\ref{fig:fig4_DFFK} for $ K_L \to \pi^0 \nu \bar\nu $, where we show the correlation with $ R^\ast_{\nu \nu} $.

\begin{figure}
\begin{center}
\includegraphics[scale=0.65]{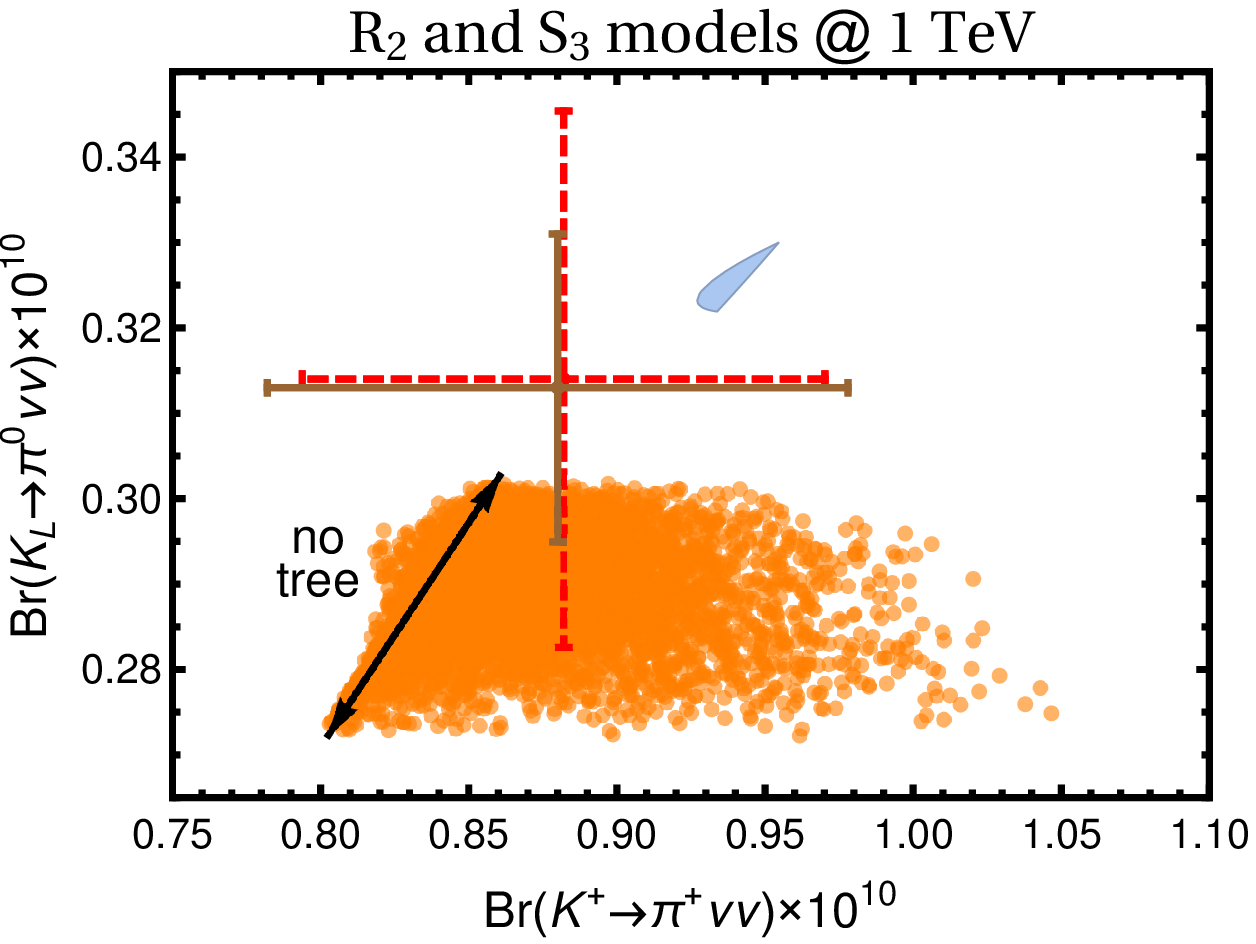}
\end{center}
\caption{\it A possible way to distinguish different NP scenarios is provided by the combined measurement of $ K \to \pi \nu \bar\nu $ rates. We indicate in blue the predictions for the $ R_2 $ model. The black line indicates the particular case for the $ S_3 $ model with $ y^{(0)}_{d \mu} = 0 $ (i.e., with no contribution already at the tree-level). In orange we give the predictions for the Scenario II of the $ S_3 $ model when $ y^{(0)}_{d \mu} $ is allowed to vary within the range $ [-1, 1] \times 10^{-4} $, taking into account the radiative corrections discussed in the main text.}
\end{figure}

\subsection{Comparison of models}
Regarding correlations with the $ B $ sector, we note that in the $ R_2 $ model we have the same couplings for the processes $ b \to s \mu^+ \mu^- $ and $ s \to d \nu \bar\nu $. This is due to the requirement that Yukawa couplings of $R_2$ to tau neutrinos are suppressed in order to enhance couplings to muons and thus the neutrinos have muonic flavors. In the $ S_3 $ model, the correlation among $ b \to s \mu^+ \mu^- $ and $ s \to d \nu \bar\nu $ depends on the way corrections to rare kaon decays are generated: when they first arrive at the tree-level as discussed above, a correlation shows up due to the coupling of $ S_3 $ to the strange-quark present in both processes; in the case rare kaon decays are generated at one-loop level, the main contribution is due to the couplings to taus and the correlation with $ b \to s \ell^+ \ell^- $ is not significant, but instead a correlation with $ b \to s \nu \bar\nu $ is relevant.

We now shift to a short discussion of the model in \cite{Crivellin:2017zlb}, which contains the triplet $ S_3 $ together with a weak-singlet $ S_1 = (\mathbf{\bar{3}}, \mathbf{1}, 1/3) $. By construction, in \cite{Crivellin:2017zlb} there is no contribution at tree-level to $ s \to d \nu \bar\nu $. This is not valid anymore after one-loop corrections are taken into account. For $ S_3 $, we have exactly the same contribution as in our present case, i.e., the same one-loop generated coupling of $ S_3 $ to a down-quark and a neutrino. For $ S_1 $, the discussion is very similar: the relevant topologies are the vertex diagram called ``V'' on Fig.~\ref{fig:diagrams} and the self-energy diagram while the diagram with $ W $ boson coupling to $S_1$ is absent. Here as well, we have checked that our calculation is gauge invariant (see Appendix~\ref{app:loopfunctions} for more details). With the numerical values found in~\cite{Crivellin:2017zlb} and both LQ masses degenerate at 1~TeV, $ K^\pm \to \pi^\pm \nu \bar\nu $ gets suppressed by $ \approx 24\% $, while $ K_L \to \pi^0 \nu \bar\nu $ gets suppressed by $ \approx 34\% $ with respect to the SM prediction. These are larger effects with respect to the case $ S_3 $ treated above due to the large values of $ y_{b \tau}, y_{s \tau} $ found in~\cite{Crivellin:2017zlb}.

\subsection{Comment on the process $ K \to \pi \mu^+ \mu^- $}

In Ref. \cite{Crivellin:2016vjc} the authors noticed that  four-fermion
operators that can explain the $ B $-physics anomalies leave their imprint also on the analoguos $ s \to d \ell^+ \ell^- $  and $ s\bar d \to \ell^+ \ell^- $ transitions. In the SM, the process $ K^+ \to \pi^+ \ell^+ \ell^- $ is dominated by long-distance effects. Despite that, small differences in the electron with respect to the muon channel could provide additional evidence for LFUV NP. However, the NP effects discussed here provide modulations below actual experimental errors. See Appendix~\ref{sec:commentonKtopiellell} for further details.

\section{Conclusion}\label{sec:concl}

In case the Lepton Flavor Universality violating observables in $ B $-physics ---$ R_{D^{(\ast)}} $, $ R_{K^{(\ast)}} $---are confirmed as true New Physics messengers, it becomes of fundamental importance to look for their signatures in other low-energy processes, in order to unveil their flavor structure. In this respect, the rare kaon decays $ K \to \pi \nu \bar\nu $, in both charged and neutral channels, are very clean theoretical probes of New Physics. Moreover, in the coming years the measurement of their branching ratios is envisaged with $ \sim 10\% $ precision.

We have considered two scenarios of scalar leptoquarks which explain the $R_{K^{(*)}}$  anomaly. In the first scenario, the weak-doublet leptoquark $R_2$ contributes to $B \to K^{(*)} \mu^+ \mu^-$ decay amplitude at loop level. In the second scenario the weak triplet leptoquark $S_3$ contributes to $B \to K^{(*)} \mu^+ \mu^-$ at tree level.

In the scenario with $R_2$ leptoquark, new contributions to
$s\to d\nu\bar\nu$ transitions are also radiatively suppressed and
imply mild corrections of the order $10\%$ relative to the Standard
Model predictions for the branching ratios of $K \to \pi \nu\bar\nu$.
In scenario with $S_3$ leptoquark, tree-level contributions to
$ s \to d \nu \bar\nu $ transitions are in principle possible and may
largely enhance the $K \to \pi \nu\bar\nu$ rates. In case the
tree-level $S_3$ contributions are set to zero, the $ 1/\lambda^2 $
Cabibbo-enhanced loop contributions could provide large enough shifts
in $ K \to \pi \nu \bar\nu $ to be observed in future experiments.
Furthermore, when we allow for tree-level Yukawas of $S_3$ to down
quarks, sizable enhancement of the branching ratio for
$ K^\pm \to \pi^\pm \nu \bar\nu $ cannot be excluded.
More generally, the combined
measurement of $ K^\pm \to \pi^\pm \nu \bar\nu $ and
$ K_L \to \pi^0 \nu \bar\nu $ may provide evidence in favor of one or
the other model. This discussion may be generalized to other
leptoquark states, as it has been briefly pointed out for the
singlet+triplet model of Ref.~\cite{Crivellin:2017zlb}.  In addition
to the leptoquark effects on the branching ratios for both decays
$ K^\pm \to \pi^\pm \nu \bar\nu $ and $ K_L \to \pi^0 \nu \bar\nu $ ,
we suggest to make correlation between the branching ratios of these
decays and the ratios $R_K$ and $R_{K^\ast}$, as well as with
$R_{\nu \nu}$.

The setup presented in this work could help to distinguish among leptoquark scenarios,  designed to resolve the $ B $-meson anomalies.  
The future results of both NA62 and KOTO experiments will shed more light on understanding the nature of the observed lepton flavor universality violation.

\section*{Acknowledgments}
We acknowledge support of the Slovenian Research Agency through research core funding No. P1-0035.
We thank Olcyr de Lima Sumensari for useful communication, and Ilja Dor\v{s}ner, Darius Alexander Faroughy and Jos\'{e} Ocariz for discussions.

\appendix

\section{Numerical values}\label{app:numValues}

We consider the following numerical values \cite{Patrignani:2016xqp,CKMfitter}
\begin{align}
m_t &= 165.95 \pm 0.73 \; {\rm GeV} \, , &  m_c &= 1.286 \pm 0.042 \; {\rm GeV} \, , \nonumber\\
m_\tau &= 1.777 \; {\rm GeV} \, , &  m_\mu &= 0.106 \; {\rm GeV} \, , \nonumber\\
M_W &= 80.385 \; {\rm GeV} \, , & s_W^2 &= 0.231 \, , \\
A &= 0.8227^{+0.0066}_{-0.0136} \, , & \lambda &= 0.22543^{+0.00042}_{-0.00031} \, , \nonumber\\
\bar\rho &= 0.151^{+0.012}_{-0.006} \, , & \bar\eta &= 0.354^{+0.007}_{-0.008} \, . \nonumber
\end{align}
In $B$-meson decay calculations we employ $ \alpha = 1/133 $, while for kaon decays we use $ \alpha = 1/137 $. At the $ M_Z $ scale we employ $ \alpha = 1/127.9 $. For the strong coupling, the central value is $ \alpha_s (M_Z)= 0.1185 $.

\section{Loop-functions}
\label{app:loopfunctions}

\subsection{Box diagrams}
Considering the box diagram with both $ W^\pm $ and $ S^{\pm 2/3}_3 $, we have the Lorentz structure $ (\gamma^\mu P_L) \otimes (\gamma_\mu P_L) $, where quark and lepton currents have been factorized through Fierz identities, external momenta have been neglected (they correspond to higher dimensional operators), and $ P_L = (1 - \gamma_5)/2 $. The flavor structure is described by the following factor:
\begin{equation}
\sum_{i,j} V^\ast_{i d} V_{j s} (V y^\ast)_{i \ell_1} (V^\ast y)_{j \ell_2} f_{i j}\,,
\end{equation}
where $ i, j $ run over $ u, c, t $, furthermore $ \ell_1 , \ell_2 $ denote two leptonic flavors (the interaction Lagrangian $ \mathcal{L}_{S_3} $ is given in the flavor basis for neutrinos), and $ f_{i j} $ is a loop-function. Using the unitarity of the CKM matrix, the up-quark contributions can be ``absorbed" into the distinct contributions labeled as $ \{ c,c \}, \{ c,t \}, \{ t,c \}, \{ t,t \} $. We then have, instead of the previous expression:
\begin{equation}
\sum_{i,j \neq u} V^\ast_{i d} V_{j s} (V y^\ast)_{i \ell_1} (V^\ast y)_{j \ell_2} (f_{i j} - f_{i u} - f_{u j} + f_{u u}) \, .
\end{equation}

In the unitary gauge it turns out that the longitudinal-component of the $ W $ propagator contributing to the loop-function $ f_{i j} $ diverges, but the flavor-blind divergent part vanishes in the combination $ \tilde{f}_{i j} \equiv f_{i j} - f_{i u} - f_{u j} + f_{u u}$. The finite loop-function $ \tilde{f}_{i j}$ is given by
\begin{eqnarray}\label{eq:boxTripletLQ}
&& \tilde{f}_{c c} = - 2 \frac{x_c}{x_S} \, , \\
&& \tilde{f}_{c t} = \tilde{f}_{t c} = \frac{x_c}{x_S} \frac{2 (-1 + x_t) \log (x_c) + (2 + x_t) \log (x_t)}{(-1 + x_t)} \, , \nonumber\\
&& \tilde{f}_{t t} = \frac{x_t}{x_S} \frac{-2 + x_t + x_t^2 - 3 x_t \log (x_t)}{(-1 + x_t)^2} \, , \nonumber
\end{eqnarray}
after expansion over small $ x_c $ and large $ x_S $, where $ x_i = m_i^2 / M_W^2 $, $ i = c, t $, and $ x_S = M_{S_3}^2 / M_W^2 $.

The correspondent contributions in the case of the $ R_2 $ model are
\begin{eqnarray}
&& F_{c c} = - 4 \frac{x_c}{x_{R_2}} (1 + \log (x_c) - \log (x_{R_2}) / 4) \, , \\
&& F_{c t} = F_{t c} = \sqrt{x_c x_t} \frac{(- 4 + x_t) (- 1 + x_{R_2}) \log (x_t) - (- 1 + x_t) (- 4 + x_{R_2}) \log (x_{R_2})}{(-1 + x_t) (x_t - x_{R_2}) (- 1 + x_{R_2})} \, , \nonumber\\
&& F_{t t} = - \frac{x_t}{x_{R_2}} \frac{(4 + (- 2 + x_t) x_t) \log (x_t) + (- 1 + x_t) x_t (4 - x_t + (- 1 + x_t) \log (x_{R_2}))}{(-1 + x_t)^2} \, , \nonumber
\end{eqnarray}
showing $ \log (x_c) $, $ x_{R_2} / \sqrt{x_c} $ and $ \log (x_{R_2}) $ enhancements compared to the previous case, Eq.~\eqref{eq:boxTripletLQ}.


\subsection{Self-energy and vertex diagrams}

Out of all possible topologies, only (SE), (V) and (VT) have equal number of powers on the Cabibbo angle $\lambda$ and $ g $. Therefore, this must be a closed set of diagrams invariant under $SU(2)_L$. It turns out that for the unitary gauge the self-energy diagram (SE) vanishes. With the same choice of gauge-fixing, the vertex topologies diverge and require renormalization. Due to weak radiative corrections the physical Yukawa couplings of $S_3$ to the $d$ quark and a lepton, $y_{d \ell}$, will be different from zero beyond tree-level. Note that for phenomenological applications we only study radiative corrections to $y_{d\ell}$ which is assumed to be zero at tree-level, in contrast to $y_{s\ell}$ and $y_{b\ell}$, which are already finite, and relatively very large, at tree-level.


%
%

We now depict the calculation of the function $ \Gamma^{\rm ren} $ in Eq.~\eqref{LagrangianSMLQM}, following the renormalization procedure discussed for instance in \cite{Basecq:1985cr}. We define the one-loop $ \bar{d}^C_L \nu^j_L S^{1/3}_3 $ function $ \Gamma $ as the sum of the different contributions $ \Gamma \equiv \Gamma_{\rm (SE)} + \Gamma_{\rm (V)} + \Gamma_{\rm (VT)} $ ($ \Gamma_{\rm (SE)} $ vanishes in the unitary gauge as previously stated). After ``absorbing" the up-quark contributions,

\begin{equation}
\Gamma_j (q^2) \equiv \Gamma_j (m_\alpha^2, q^2) - \Gamma_j (m_u^2, q^2) \, , \quad \alpha = c, t \, ,
\end{equation}




\noindent
we still get a divergent term, contrarily to the case of the box diagrams.
The counter-term is proportional to $ \bar{d}^c_L \nu^j_L S^{1/3}_3 $ (or its Hermitian conjugate), $ j = \mu, \tau $, which, note, is not present in the bare Lagrangian for our phenomenological choice $ y^{(0)}_{d j} = 0 $ (``$ (0) $" indicating the bare coupling), but is not protected under radiative corrections. 

To carry out the renormalization procedure, we let free the momenta of $ d, \nu^j, S^{1/3}_3 $, and consider throughout the calculation $ p_d^2 = p_\nu^2 = m^2_d = m^2_\nu = 0 $, where $ p_d, p_\nu $ are the four-momenta of the down-quark and neutrino, respectively, and $ p_d \cdot p_\nu = - q^2 / 2 $, where $ q = p_d - p_\nu $ is the four-momentum of the $ S^{1/3}_3 $. The subtraction is made on-shell, i.e.,

\begin{equation}\label{eq:R1subtraction}
\Gamma^{\rm ren}_j (q^2; M^2_{S_3}) = \Gamma_j (q^2) - \Gamma_j (M^2_{S_3})
\end{equation}
Note that, when calculating the NP contribution to $ K \to \pi \nu \bar\nu $ amplitudes, we will be interested by the function $ \Gamma^{\rm ren} $ calculated at $ q^2 = 0 $, which is equivalent to neglecting higher-order operators in the low-energy effective Lagrangian.

Employing the on-shell renormalization of Eq.~\eqref{eq:R1subtraction}, and neglecting here the mass of the charged lepton in the loop for simplification (kept in our phenomenological studies), we get at $ q^2 = 0 $
\begin{eqnarray}
&& \Gamma^{\rm ren}_{\ell_1} (0; M^2_{S_3}) = (-1) \frac{g^2 V^\ast_{\alpha d} (V y^\ast)_{\alpha \ell_1}}{16 \pi^2} M^2_{S_3} \Bigg[ C_{0(0, M^2_{S_3}, 0, {} M_W^2, 0, 0)} - \frac{m_\alpha^2}{M^2_{S_3}} C_{0(0, 0, 0, {} M_W^2, 0, m_\alpha^2)} \nonumber\\
&& - \left( 1 - \frac{m_\alpha^2}{M^2_{S_3}} \right) C_{0(0, M^2_{S_3}, 0, {} M_W^2, 0, m_\alpha^2)} + (- 2 C_{0(0, 0, 0, {} M_W^2, M^2_{S_3}, 0)} - \left( 2 - \frac{m_\alpha^2}{M^2_{S_3}} \right) C_{0(M^2_{S_3}, 0, 0, {} M_W^2, M^2_{S_3}, m_\alpha^2)} \nonumber\\
&& - \left( x_\alpha + \left( \frac{m_\alpha^2}{M^2_{S_3}} - 2 \right) \right) C_{0(0, 0, 0, {} M_W^2, M^2_{S_3}, m_\alpha^2)} + 2 C_{0(M^2_{S_3}, 0, 0, {} M_W^2, M^2_{S_3}, 0)}) \Bigg] + \mathcal{O} (m_{\ell_1}^2) \, ,
\end{eqnarray}

\noindent
where $ x_\alpha = m_\alpha^2 / M_W^2 $. The function $ C_0 $ has an analytical expression in terms of di-logarithms, see e.g. \cite{vanOldenborgh:1989wn}. In some cases, it has a simple expression, for instance

\begin{equation}
C_{0(0, 0, 0, {} M_W^2, M^2_{S_3}, m_\alpha^2)} = \frac{x_\alpha \log (m_\alpha^2 / M^2_{S_3}) - \log (M_W^2 / M^2_{S_3})}{(x_\alpha - 1) M^2_{S_3}} + \mathcal{O}(1/M^4_{S_3}) \, ,
\end{equation}
\noindent
showing that large logarithmic enhancements are present. We use the package \textsc{LoopTools} \cite{Hahn:1998yk} in order to generate the numerical values of the function $ C_0 $, whose arguments are given in subscript.

\subsubsection{Gauge invariance}
We resume a few elements necessary for checking explicitly the gauge invariance of the one-loop coupling of the $ S_3^{1/3} $ ($ S_3^{4/3} $) to the down-quark and a neutrino (respectively, a charged lepton), i.e., that for any gauge $ \xi_W $ the resulting one-loop expressions are independent on the choice of $ \xi_W $.
Note in particular that the unitarity of the CKM matrix eliminates terms that do not depend on the flavor of the up-type quark in the loop. It is appropriate to mention that it is easier to prove gauge invariance after the on-shell subtraction discussed in the previous Section, though of course gauge invariance holds here at all steps of the renormalization procedure (for comments that also apply here related to gauge-invariance and the on-shell subtraction, see \cite{Basecq:1985cr,GagyiPalffy:1997hh}).

The couplings of the scalar LQs to the Goldstone bosons in Fig.~\ref{fig:sub1} are calculated from the scalar potential. To full generality, we have

\begin{equation}
V (\phi, S_3) \supset - \mu^2 (\phi^\dagger \phi) + \frac{\rho}{2} (\phi^\dagger \phi)^2 + \tilde{\mu}^2 {\rm Tr} \{ S_3 S_3^\dagger \} + \alpha (\phi^\dagger \phi) {\rm Tr} \{ S_3 S_3^\dagger \} + \beta (\phi^\dagger  S_3 S_3^\dagger \phi) \, ,
\end{equation}

\noindent
where we do not include terms with four LQs (cf. Appendix A in \cite{Dorsner:2016wpm}). In the potential above, the $ \beta $ term results in a mass splitting among the different LQ fields. Note that it is not equivalent to the $ \alpha $ term, since the two terms in the LHS of the following expression are not identical

\begin{equation}
{\rm Tr} \{ S_3^\dagger S_3 \phi \phi^\dagger \} + {\rm Tr} \{ S_3 S_3^\dagger \phi \phi^\dagger \} = {\rm Tr} \{ S_3^\dagger S_3 \} {\rm Tr} \{ \phi \phi^\dagger \} \, .
\end{equation}


The scalar potential gives then the coupling


\begin{equation}
\mathcal{L} \supset - \sqrt{\frac{2 \rho}{\mu^2}} ( M^2_{S^{2/3}_3} - M^2_{S^{1/3}_3} ) \Big( S_3^{1/3} S_3^{2/3} - S_3^{-1/3} S_3^{4/3} \Big) G^- + {\rm h.c.}
\end{equation}
where $ G^\pm $ is the charged Goldstone boson of the standard model, and $ M^2_{S^{2/3}_3} - M^2_{S^{1/3}_3} = M^2_{S^{1/3}_3} - M^2_{S^{4/3}_3} $, with masses at tree-level read from $ V (\phi, S_3) $ above. For simplicity, in the main text discussing our numerical evaluations we have considered masses degenerate and equal to $ M_{S_3} $.

\subsection{Expressions for $ K \to \pi \mu^+ \mu^- $}\label{sec:commentonKtopiellell}
In the case of the model with $ S_3 $ we do not have the vertex diagram ``V''. There is an additional diagram indicated in Figure~\ref{fig:sub5}, proportional to the mass of the neutrino in the loop, and therefore negligible. The pattern of $ \lambda $ suppressions, together with $ g $ and $ y $ matrix element powers is also indicated in Table~\ref{tab:flavornunu}.
To pursue the discussion, we define the following effective Lagrangian

\begin{equation}
\mathcal{L}_{{\rm eff}: s \to d\ell\ell} \supset - \frac{G_F}{\sqrt{2}} V_{us} V^\ast_{ud} \sum_{\ell=e, \mu} \left( C^{\ell \ell}_{7V} Q^{\ell \ell}_{7V} + C^{\ell \ell}_{7A} Q^{\ell \ell}_{7A} \right) + {\rm h.c.} \, ,
\end{equation}

\noindent
where

\begin{equation}
Q^{\ell \ell}_{7V} = (\bar{d} \gamma^\mu (1 - \gamma_5) s) \times (\bar\ell \gamma_\mu \ell) \, , \qquad Q^{\ell \ell}_{7A} = (\bar{d} \gamma^\mu (1 - \gamma_5) s) \times (\bar\ell \gamma_\mu \gamma_5 \ell) \, .
\end{equation}

\noindent
Allowing for $ |y^{(0)}_{d \mu}| \sim 10^{-4} $ (see Secion~\ref{sec:treeeffects}) does not result in a large contribution to $ C^{\mu \mu (NP)}_{7V} $. For completeness, we now give the expressions for one-loop contributions, taking $ y^{(0)}_{d \mu} = 0 $. While $ C^{e e (NP)}_{7V} $ in our present case vanishes at one-loop order, the expression for $ C^{\mu \mu (NP)}_{7V} $, dominated by the top contribution, is implicitly given by


\begin{eqnarray}
&& - \frac{G_F}{\sqrt{2}} V_{us} V^\ast_{ud} C^{\mu \mu (NP)}_{7V} = - \frac{g^2 V^\ast_{t d} (V y^\ast)_{t \mu} y_{s \mu}}{128 \pi^2} \\
&& \times \Bigg[ 2 C_{0(0, 0, 0, {} M_W^2, M^2_{S_3}, 0)} + \left( 2 - \frac{m_t^2}{M^2_{S_3}} \right) C_{0(M^2_{S_3}, 0, 0, {} M_W^2, M^2_{S_3}, m_t^2)} \nonumber\\
&& + \left( x_t + \left( \frac{m_t^2}{M^2_{S_3}} - 2 \right) \right) C_{0(0, 0, 0, {} M_W^2, M^2_{S_3}, m_t^2)} - 2 C_{0(M^2_{S_3}, 0, 0, {} M_W^2, M^2_{S_3}, 0)} \Bigg] \nonumber
\end{eqnarray}

\noindent
which is numerically given by
\begin{equation}
C^{\mu \mu (NP)}_{7V} = \left[ - y_{s \mu} y_{b \mu} (1.7 + 0.7 i) + y_{s \mu}^2 (0.07 + 0.03 i) \right] \times 10^{-5} \, ,
\end{equation}
for $ M_{S_3} = 1 $~TeV, where complex phases are {\bf CP}-odd, and where only the top contribution coming from the vertex and self-energy topologies are kept (superscripts ``$ (0) $" have been dropped for readability). Given the smallness of $ y_{s \mu} y_{b \mu} $ and $ y_{s \mu}^2 $ as extracted from the global fit, together with their overall small coefficients, NP effects discussed here are largely suppressed as stated in the main text. Similar conclusions also hold for the process $ K_L \to \ell^+ \ell^- $, which by an analogous reasoning probes directly $ C^{\mu \mu (NP)}_{7A} $, equals to $ - C^{\mu \mu (NP)}_{7V} $. Given the much milder variation in the case of the $ R_2 $ model, similar conclusions also hold.

\input{full_bib_3.tex}


\end{document}

%% file: Diagrams.tex
\begin{figure}
\centering
\begin{subfigure}{.35\textwidth}
\centering
	%
	%
%
\SetScale{1}\SetWidth{1.2}\begin{picture}(100,100)(0,-10)
	\ArrowLine(0,60)(27,60)
	\ArrowLine(27,60)(67,60)
	\ArrowLine(94,60)(67,60)
	\Text(13,48)[b]{ {\color{black} $ s $ } }
	\Text(47,48)[b]{ {\color{black} $ u' $ } }
	\Text(80,47)[b]{ {\color{black} $ \nu_{\ell_2} $ } }
	\ArrowLine(27,0)(0,0)
	\ArrowLine(67,0)(27,0)
	\ArrowLine(67,0)(94,0)
	\Text(13,12)[t]{ {\color{black} $ d $ } }
	\Text(47,13)[t]{ {\color{black} $ u'' $ } }
	\Text(80,11)[t]{ {\color{black} $ \nu_{\ell_1} $ } }
	\Photon(27,0)(27,60){4}{6}
	\DashLine(67,0)(67,60){3}
	%
	%
	\Text(15,30)[c]{ {\color{black} $ W $ } }
	\Text(90,30)[c]{ {\color{black} $ S_3^{2/3} $ } }
\end{picture}
  \caption{\it Box diagram (Box).}
  \label{fig:sub4}
\end{subfigure}
\begin{subfigure}{.6\textwidth}
\centering
\SetScale{1}\SetWidth{1.2}\begin{picture}(100,100)(0,-10)
	\ArrowLine(0,60)(20,60)
	\ArrowLine(60,60)(20,60)
	\ArrowLine(80,60)(60,60)
	\Text(10,48)[b]{ {\color{black} $ s $ } }
	\Text(44,48)[b]{ {\color{black} $ \nu_\ell $ } }
	\Text(77,47)[b]{ {\color{black} $ \ell $ } }
	\ArrowLine(20,0)(0,0)
	\ArrowLine(60,0)(20,0)
	\ArrowLine(60,0)(80,0)
	\Text(10,12)[t]{ {\color{black} $ d $ } }
	\Text(44,13)[t]{ {\color{black} $ u' $ } }
	\Text(77,11)[t]{ {\color{black} $ \ell $ } }
	\Photon(20,0)(38,29){4}{3}
	\Photon(42,31)(60,60){4}{3}
	\DashLine(60,0)(20,60){3}
	\Text(19,24)[c]{ {\color{black} $ W $ } }
	\Text(70,24)[c]{ {\color{black} $ S_3^{1/3} $ } }
\end{picture}
  \caption{\it Crossed box diagram, relevant for $s \to d \ell^+
    \ell^-$ (Box')}
  \label{fig:sub5}
\end{subfigure}
\begin{subfigure}{.3\textwidth}
\centering
\SetScale{1}\SetWidth{1.2}\begin{picture}(100,100)(0,-20)
	\ArrowLine(0,60)(40,60)
	\ArrowLine(80,60)(40,60)
	\Text(20,50)[b]{ {\color{black} $ s $ } }
	\Text(60,50)[b]{ {\color{black} $ \nu_{\ell_2} $ } }
	\ArrowLine(27,0)(0,0)
	\ArrowLine(53,0)(27,0)
	\ArrowLine(53,0)(80,0)
	\Text(13,12)[t]{ {\color{black} $ d $ } }
	\Text(73,10)[t]{ {\color{black} $ \nu_{\ell_1} $ } }
	\Text(40,-13)[b]{ {\color{black} $ u' $ } }
	\DashLine(40,30)(40,60){3}
	\Photon(27,0)(40,30){3}{4}
	\DashLine(53,0)(40,30){3}
	\Text(28,20)[c]{ {\color{black} $ W $ } }
	\Text(30,40)[c]{ {\color{black} $ S^{1/3}_3 $ } }
	\Text(63,20)[c]{ {\color{black} $ S^{2/3}_3 $ } }
\end{picture}
  \caption{\it Vertex--like diagram (VT).}
  \label{fig:sub1}
\end{subfigure}
\begin{subfigure}{.3\textwidth}
\centering
\SetScale{1}\SetWidth{1.2}\begin{picture}(100,100)(0,-20)
	\ArrowLine(0,60)(40,60)
	\ArrowLine(80,60)(40,60)
	\Text(20,50)[b]{ {\color{black} $ s $ } }
	\Text(60,50)[b]{ {\color{black} $ \nu_{\ell_2} $ } }
	\ArrowLine(20,0)(0,0)
	\ArrowLine(40,0)(20,0)
	\ArrowLine(40,0)(60,0)
	\ArrowLine(60,0)(80,0)
	\Text(10,12)[t]{ {\color{black} $ d $ } }
	\Text(70,10)[t]{ {\color{black} $ \nu_{\ell_1} $ } }
	\Text(30,12)[t]{ {\color{black} $ u' $ } }
	\Text(50,12)[t]{ {\color{black} $ {\ell_1} $ } }
	\DashLine(40,0)(40,60){3}
	\PhotonArc(40,0)(20,180,0){3}{6}
	\Text(10,-15)[b]{ {\color{black} $ W $ } }
	\Text(30,30)[c]{ {\color{black} $ S^{1/3}_3 $ } }
\end{picture}
  \caption{\it Vertex--like diagram (V).}
  \label{fig:sub2}
\end{subfigure}
\begin{subfigure}{.35\textwidth}
\centering
\SetScale{1}\SetWidth{1.2}\begin{picture}(100,100)(0,-20)
	\ArrowLine(0,60)(40,60)
	\ArrowLine(80,60)(40,60)
	\Text(20,50)[b]{ {\color{black} $ s $ } }
	\Text(60,50)[b]{ {\color{black} $ \nu_{\ell_2} $ } }
	\Line(11,0)(0,0)
	\ArrowLine(31,0)(11,0)
	\Line(40,0)(31,0)
	\ArrowLine(40,0)(80,0)
	\Text(5,12)[t]{ {\color{black} $ d $ } }
	\Text(60,10)[t]{ {\color{black} $ \nu_{\ell_1} $ } }
	\Text(21,5)[b]{ {\color{black} $ u' $ } }
	\PhotonArc(21,0)(10,180,0){3}{4}
	\DashLine(40,0)(40,60){3}
	\Text(42,-7)[t]{ {\color{black} $ W $ } }
	\Text(30,30)[c]{ {\color{black} $ S^{1/3}_3 $ } }
\end{picture}
  \caption{\it Self-energy--like diagram (SE).}
  \label{fig:sub3}
\end{subfigure}
\caption{\it Possible diagrams contributing to
  $ K \to \pi \nu \bar\nu$ or $K \to \pi \ell^+ \ell^-$ for the $S_3$
  LQ model when $ y_{d \mu} = 0 $ (In the $ R_2 $ model, only the box diagrams
  in the first line contribute via the $R_2^{2/3}$ state. In this model,
  the lepton lines in the diagram should be reversed.).  
A similar set of diagrams is found when replacing the
  $ W^\pm $ gauge boson by the respective Goldstone bosons.}
\label{fig:diagrams}
\end{figure}

%% file: full_bib_3.tex
\newpage{\pagestyle{empty}\cleardoublepage}